# Current Unsolved Problems in Planetary Nebulae Research

**Sun Kwok** [1,2,*], **Bruce Balick** [3], **You-Hua Chu** [4], **Bruce J. Hrivnak** [5], **Alberto López** [6], **Quentin Parker** [2], **Raghvendra Sahai** [7] **and Albert Zijlstra** [8]

1. Department of Earth, Ocean, and Atmospheric Sciences, University of British Columbia, Vancouver, BC V6T 1Z4, Canada
2. Laboratory for Space Research, University of Hong Kong, Hong Kong, China; quentinp@hku.hk
3. Department of Astronomy, University of Washington, Seattle, WA 98195, USA; balick@uw.edu
4. Department of Physics, National Sun Yat Sen University, Kaohsiung 80424, Taiwan; yhchu@asiaa.sinica.edu.tw
5. Department of Physics and Astronomy, Valpariso University, Valpariso, IN 46383, USA; bruce.hrivnak@valpo.edu
6. Instituto de Astronomía, Universidad Nacional Autónoma de México, Campus Ensenada, Ensenada 22760, Baja California, Mexico; jal@astro.unam.mx
7. Jet Propulsion Laboratory, California Institute of Technology, Pasadena, CA 91109, USA; sahai@jpl.nasa.gov
8. Jodrell Bank Centre for Astrophysics, Department of Physics & Astronomy, University of Manchester, Manchester M13 9PL, UK; albert.zijlstra@manchester.ac.uk
* Correspondence: sun.kwok@ubc.ca or sunkwok@hku.hk; Tel.: +1-778-858-5752

## Abstract

While there has been significant progress in our understanding of the origin and evolution of planetary nebulae in the last 50 years, there remain several unsolved problems. These include the true 3D morphological structure of the nebulae, origin of multipolar nebulae, the dust and molecular distribution relative to the optical nebulosity, large-scale structures outside of the main nebulae, the relevance of binarity to planetary nebulae evolution, and a precise definition of the planetary nebula phenomenon. The long-standing problem of elemental abundance discrepancy still remains unsolved. In this paper, we summarize current observations related to these problems and present possible future directions to tackle them.

## 1. Introduction

Although planetary nebulae have been known for over 200 years, the journey to understand the nature and origins of this astronomical phenomenon has been a long one. The observation of emission lines in NGC 6543 by William Huggins in 1864 established that planetary nebulae are composed of gases and not a collection of stars. In the early 20th century, Curtis determined from the galactic distribution of planetary nebulae that they belong to the old star population and are not young stars as they were previously believed [1]. The evolutionary stage of planetary nebulae as descendants of red giants and precursors of white dwarfs was proposed by Shklovsky in 1956 [2]. Assuming that the luminosity of planetary nebulae is provided by hydrogen-shell burning on an electron-degenerate carbon–oxygen core of the progenitor asymptotic giant branch (AGB) star, Paczynski showed that planetary nebulae evolve horizontally across the Hertzsprung–Russell diagram [3]. His theoretical tracks were contrary to the observed distribution of planetary nebulae at the time. After the observations were later found to be erroneous, the Paczynski tracks became the accepted model of planetary nebulae central-star evolution, leading to a quantitative understanding of stellar evolution between red giants and white dwarfs.

Although the nebular shell was long believed to be the result of a sudden ejection of the hydrogen envelope of the AGB star, none of the proposed mechanisms were able to satisfactorily explain the observations. The discovery of mass loss from AGB stars led to the proposal that the nebular shell is formed by the compression of AGB stellar wind by a later-developed fast wind from the central star [4,5]. Integrated models of central-star evolution coupled with interacting wind dynamics have successfully reproduced many of the morphological structures of planetary nebulae [6]. The standard model that emerged was of planetary nebulae forming from the gradual post-AGB evolution of an evolved star and the corresponding expansion of the surrounding material ejected during the AGB phase, with the planetary nebulae phase beginning when the star becomes hot enough (>25,000 K) to photoionize the nebula. Objects in this transition phase between the mass-losing AGB star and the photoionization of the nebula are termed proto-planetary nebulae.

As observational techniques continue to improve and extend to all wavelengths across the electromagnetic spectrum, new problems have emerged. Space-based optical telescopes have revealed faint and extended structures, and infrared and millimeter-wave observations have found the presence of molecular and solid-state components in addition to the ionized gas component of planetary nebulae. The state of our understanding of the planetary nebula phenomenon at the end of the 20th century was well summarized by two monographs [7,8]. Updates on the current state of the field are given by two recent reviews [9,10]. In this paper, we highlight observed phenomena that do not fit easily into this standard model of the formation of planetary nebulae and list some of the currently unsolved problems in planetary nebulae research.

## 2. What Is a Planetary Nebula?

The question of "what is a planetary nebula" is not as straightforward as one might think. Many different phenomenological object types in the universe (emission-line galaxies, reflection nebulae, H II regions, Wolf–Rayet nebulae, novae, supernova remnants, young stellar objects, and symbiotic stars) can closely mimic planetary nebulae in their observable characteristics [11]. The confusion between planetary nebulae and other objects has been a major problem in early planetary nebulae catalogs [12,13].

The most recent planetary nebulae catalog, the Hong Kong/AAO/Strasbourg Planetary Nebulae Database (HASH) [14], has made major progress in clarifying this issue and in removing suspect objects. The HASH project has the most complete inventory for galactic and Magellanic Cloud planetary nebulae and has doubled the total galactic planetary nebulae known and compiled over the previous 200 years. Their procedure of identifying planetary nebulae is described in reference [15]. The HASH catalog has led to significant numbers of reclassifications in SIMBAD (the Set of Identifications, Measurements and Bibliography for Astronomical Data) from planetary nebulae to other types of contaminating compact and resolved narrow emission-line sources.

It is now clear that observational properties alone cannot uniquely define what is a planetary nebula, and a combination of observational and evolutionary properties is needed. A possible definition of a planetary nebula is the following: an emission-line object showing an ionized circumstellar shell with some degree of symmetry surrounding a hot, compact star evolving from the AGB to white dwarf stage [10,16]. Furthermore, the ionized material itself has to have come from the star that is doing the ionization—i.e., it is not donated material from a companion. A decision tree to assist this process has been published [15].

## 3. The Distance Problem

One of the long-standing problems of planetary nebulae was individual distances. Distances are important in determination of luminosities and the placement of central stars of planetary nebulae on the Herztsprung–Russell diagram. Distances are also needed to convert angular sizes to physical sizes and for extracting expansion ages from measurements of Doppler shifts [17].

Statistical distances were used for many years, but they were not reliable for individual objects. The recent *Gaia* distances have greatly improved the situation. For planet nebulae within 2–3 kpc and central stars with apparent visual magnitudes less than 20, accurate distances can be obtained. However, *Gaia* distances depend on the correct identification of the central stars. If a central star is on the cooling track or suffers from significant foreground extinction, it may be too faint for *Gaia*. It is estimated that ~20% of the central stars in recent distance determinations [18,19] may be misidentified [20]. For extended planetary nebulae with angular sizes larger than 1 arcmin, the situation is better as the central stars are easier to identify [21]. Further improvement in distances is expected in future *Gaia* data releases. Thus, planetary nebulae distances are therefore no longer an unsolved problem.

## 4. Morphological Structure of Planetary Nebulae

A common perception of a planetary nebula is an expanding round or elliptical shell of ionized gas surrounding a central star. However, the reality is much more complicated. After the introduction of charge-coupled device (CCD) imaging, we learned that many planetary nebulae have multiple shell structures consisting of rims, shells, crowns, and haloes [22,23]. Large, spherical haloes are found outside of the main nebula, including some of the most well-known objects such as NGC 7293 (the Helix Nebula) [24] and NGC 6543 (the Cat's Eye Nebula) [22].

A large number of planetary nebulae (e.g., Hb 5, NGC 6302) show bipolar morphology with a pair of symmetric lobes centered on the central star. Based upon imaging and spectro-kinematic observations, many well-known planetary nebulae, e.g., NGC 6720 (the Ring Nebula) [25], NGC 6853 (the Dumbbell Nebula) [26], NGC 7293 (the Helix Nebula) [27], and NGC 7027 [28,29], are also believed to be bipolar with the bipolar axis oriented at a large angle from the plane of the sky. The complications with morphological classifications based on 2D imaging are illustrated in Figure 1, where NGC 6543, NGC 7009, and NGC 2392 are shown to have similar intrinsic 3D structures although their apparent morphologies in the sky are different.

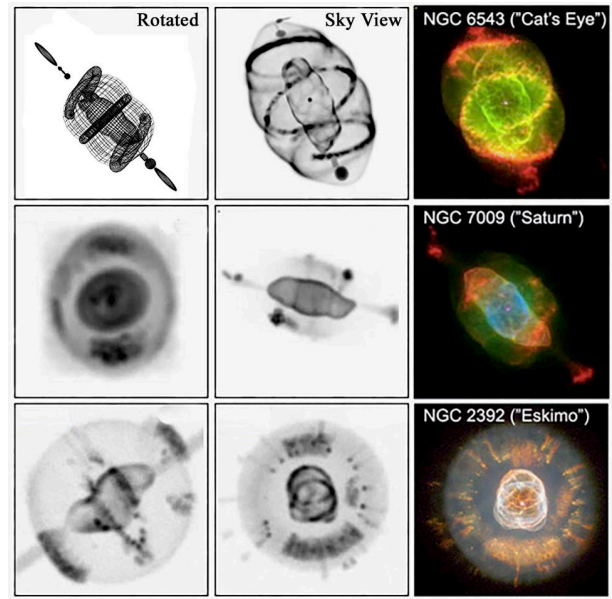

**Figure 1.** Illustration of the similar intrinsic structures between NGC 6543 (Cat's Eye Nebula), NGC 7009 (the Saturn Nebula), and NGC 2392 (the Eskimo Nebula). The *HST* images of the objects are shown in the right column. The central panels are simulated images from the SHAPE morpho-kinematic model as seen in the sky. When rotated, the simulated image of NGC6543 (**top left**) looks like NGC 7009, the simulated image of NGC 7009 (**center left**) looks like NGC 2392, and the simulated image of NGC 2392 (**bottom left**) looks like NGC 6543. The SHAPE models are adapted from references [30,31] and the images on the right column are from NASA *HST* web pages.

The bipolar lobes can have either open (e.g., NGC 6302) or closed (e.g., Hb 5) ends. Such bipolar lobes are likely to be shaped by the interaction between a fast outflow from the central star with an external medium of varying density distribution created by its AGB progenitor. One possible mechanism is outflows channeled through a de Laval nozzle confined by an external pressure gradient [32]. This model can simulate the unique hourglass shape seen in some planetary nebulae (e.g., Hb 12 [33]).

While we were still debating the fraction of planetary nebulae that are bipolar, a new class of multipolar planetary nebulae was discovered [34,35]. An example is NGC 2440, where two pairs of bipolar lobes of approximately equal lengths are observed at different orientations within a spherical envelope (Figure 2). After the launch of the *Hubble Space Telescope (HST)*, more multipolar nebulae were discovered [36–38]. Three pairs of bipolar lobes are seen in the planetary nebula M1-37 (Figure 2) and two pairs of lobes in each of IC 5117 and Hen2-447 (Figure 3). The lobes are approximately equal in size, but orientated at different angles from the plane of the sky [36]. However, in the bipolar nebula Hb 5, a pair of shorter secondary lobes can be seen emerging from the core but in different directions from the main bipolar lobes [39].

Multipolar structures may not be revealed until a deeper image is taken, or the object is observed in the infrared (Figure 4). Such multipolar structures can already be seen in proto-planetary nebulae (Figure 5), suggesting that these structures emerge early in post-AGB evolution.

Although there seems to be a large diversity of planetary nebulae morphologies, many may share a similar intrinsic 3D structure when the effects of orientation and sensitivity are taken into account [40]. With improved angular resolution and dynamic range imaging, more planetary nebulae may show multipolar structures.

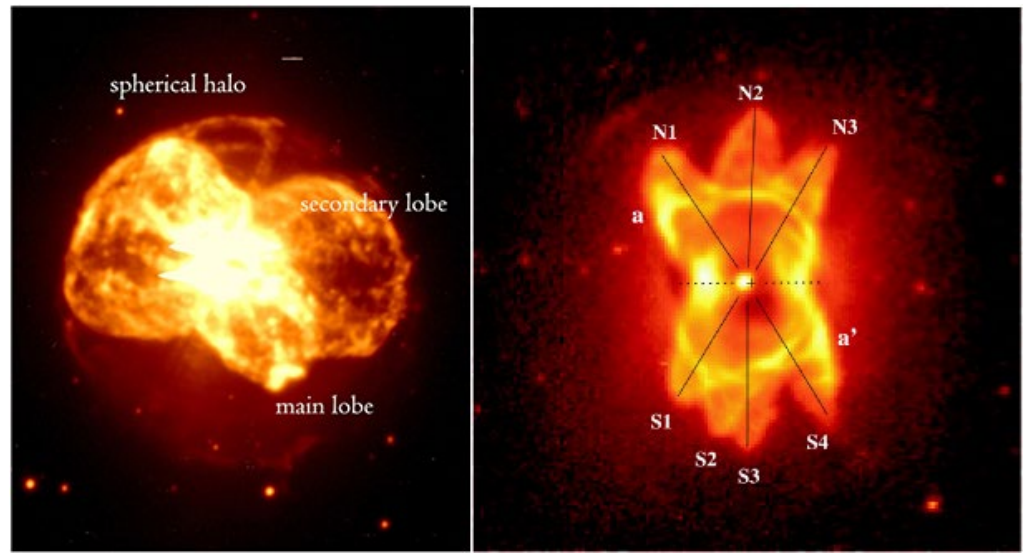

**Figure 2.** Examples of multipolar nebula. Left: NGC 2440 was discovered to be a multipolar nebula by Lopez et al. (1998) [35]. **Left**: a false-color [N II] image of NGC 2440 taken with the *Canada-France-Hawaii Telescope*. Figure adapted from reference [41]. **Right**: *HST* WFPC2 Hα image of the multipolar nebula M1-37. The very faint regions are coded in red, whereas the brighter regions are coded in orange and yellow. The dotted line shows the minor axis, and the radial vectors indicate the position angles of selected lobes. Figure adapted from reference [36].

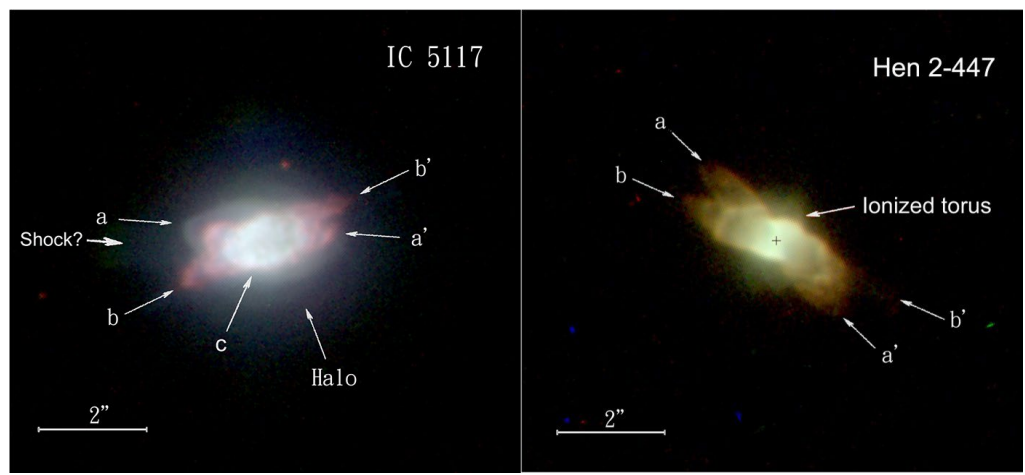

**Figure 3.** *HST* WFPC2 color composite images of IC 5117 and Hen2-447 ([O III] in blue, Hα in green, and [N II] in red). The bipolar axes are labeled as *a–a'* and *b–b'*. Figure adapted from reference [37].

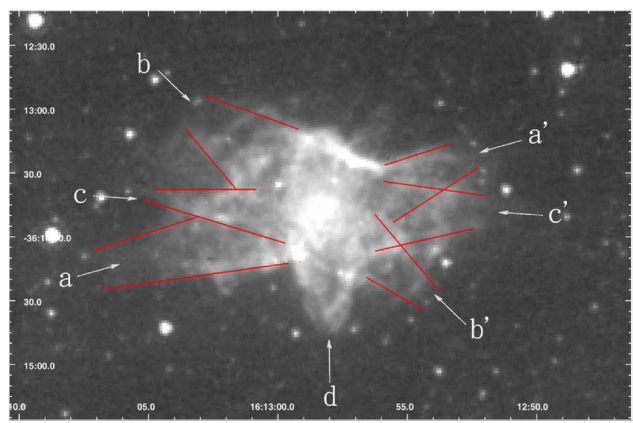

**Figure 4.** *Spitzer* IRAC 5.8 μm image of NGC 6072 with the three pairs of bipolar lobes marked as *a–a'*, *b–b'*, *c–c'*. The equatorial disk is marked as *d*. Figure adapted from reference [42].

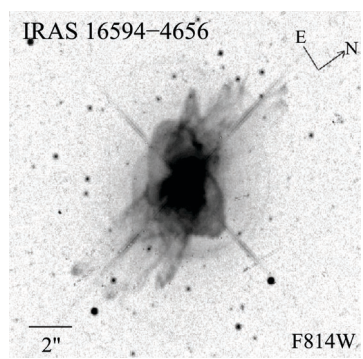

**Figure 5.** *HST* WFPC2 imaging of the proto-planetary nebula IRAS 16594−4656 showing multipolar structures. Figure adapted from reference [43].

The cause of the formation of bipolar and multipolar nebulae has been a major topic of investigation in the past several decades. Below we discuss some of the physical mechanisms that are related to these structures.

### 4.1. Wind Shaping

With the advent of CCD imaging and then especially high sensitivity, high resolution imaging with the *HST*, it became apparent that the majority of planetary nebulae are not spherical and that the morphology of the nebulae is complex. Quite often they appear as bipolar or even multipolar or point-symmetric, with haloes and jets, as discussed above. How a spherical, mass-losing AGB star could produce such morphological features became a topic of much interest [44]. We now recognize that the morphology of planetary nebulae is shaped by the interaction of stellar outflows during the AGB phase, post-AGB phase, and beyond. In the original 1D formulation of the interacting-wind model, the morphology of the nebular shell is the result of interacting wind dynamics, and the expansion of the nebular shell is driven by thermal pressure of the hot bubble created by the shocked fast wind [8]. The existence of the "hot bubble" was confirmed by X-ray observations [45–47] (Figure 6).

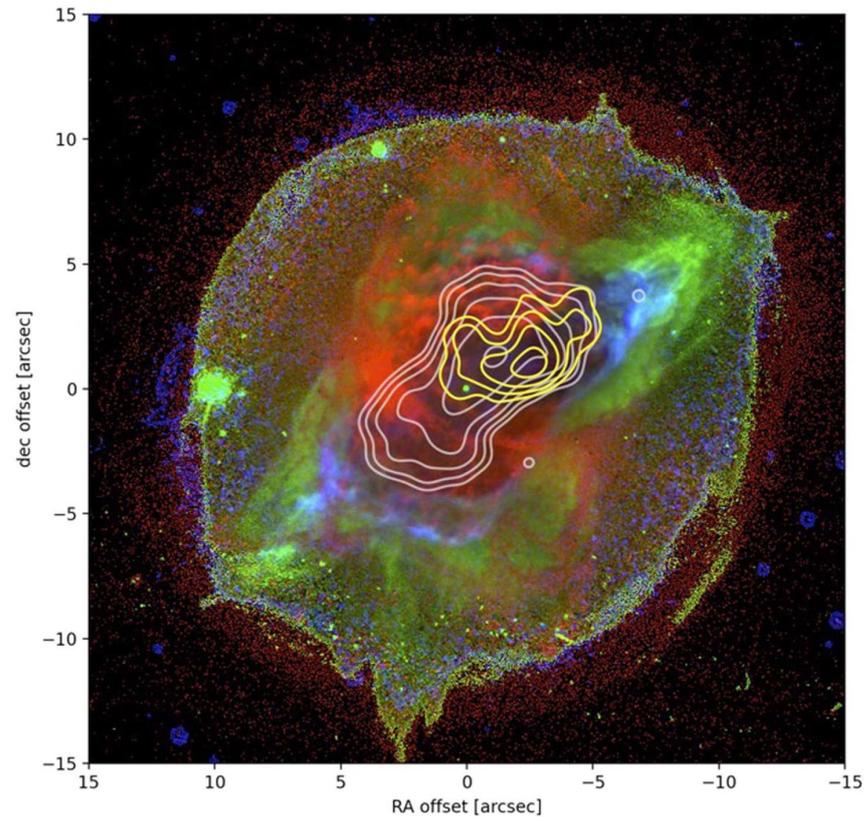

**Figure 6.** NGC 7027 is one of many planetary nebulae where a hot bubble formed by shocked fast wind has been detected in X-ray. In this image, the color composite image is from *HST* WFC3 observations of the optical nebula (red is Hα/Hβ, green is [S II]/Hα, and blue is [Fe II]). White and yellow contours trace hard and soft X-ray emissions, respectively, from data obtained with the Chandra X-ray Observatory [48]. Figure adapted from reference [49].

The emergence of asymmetric structures is likely the result of asymmetries in either the AGB wind or the later-developed fast wind [50,51]. Extensive numerical models have been developed to simulate bipolar nebulae with the interacting-wind model assuming asymmetry in the AGB wind [52–55].

What is the origin of such asymmetries and when did they first develop? Asymmetric structures have been found in OH/IR stars [56]. Recent *Atacama Large Millimeter/Submillimeter Array (ALMA)* observations have found that the seeds of planetary nebular asymmetries (equatorial density enhancements, bipolar flows and spiral structures) are already present in the AGB winds [57]. These asymmetries could be amplified into their final shape by the fast wind from the central star during the planetary nebula phase.

The clues to when morphological shaping begins can be found in proto-planetary nebulae, objects in transition between AGB and planetary nebulae phases of evolution. Searching at infrared wavelengths for objects with infrared colors between those of AGB stars and planetary nebulae led to the identification of proto-planetary nebulae [58]. Many of these objects display spectra characteristics of F-G supergiants [59]. Since these objects have not yet evolved to the high temperatures needed to photoionize the surrounding nebula, they appear smaller and are seen only faintly in scattered light. Imaging with the *HST* showed that they already possess bipolar and, in some cases, multipolar shapes, which was interpreted as indicating that much of the shaping happened early in the proto-planetary nebulae phase or even in the late-AGB phase [60–63].

The initial investigations of shaping mechanisms focused on magnetic fields and binary companions [64,65]. However, it became apparent that magnetic fields would decay too rapidly to sustain their field strength over the relevant times and any method to strengthen the fields would most likely require a binary companion [66]. Although magnetic fields may be effective especially with jet collimation [67], binary companions were deemed the likely driving mechanism for the main shaping [68]. Such companions can gravitationally modify the density distribution of the circumstellar envelope, enhancing the material density in the orbital plane to produce a circumbinary torus or disk. Also, mass transfer to a companion could lead to an accretion disk around it, which collimates a fast wind and leads to narrow-waist bipolar planetary nebulae [69]. Gravitational interactions during the binary orbit could also create the arcs, rings (Section 4.3) and spirals (Section 4.2) observed.

### 4.2. Collimated Flows

Our current understanding of planetary nebulae dynamics is complicated by the observations of collimated outflows from the central star. While the radiatively driven 2000–4000 km s$^{-1}$ fast wind from the central star [70,71] is largely spherically symmetric, collimated outflows attributed to mass ejection events, either continuous or episodic, have been observed. Collimated outflows with speeds of ~370 km s$^{-1}$ have been found in He2-111 [72], far exceeding the expansion velocities of 20–50 km s$^{-1}$ commonly observed in the main shell of planetary nebulae. Similarly high-velocity outflows (850 km s$^{-1}$ and 500 km s$^{-1}$) have also been found in IRAS 17423–1755 and in MyCn 18, respectively [73,74].

The mechanisms responsible for the creation of these collimated outflows are not known. Possible theories include the extraction of momentum and energy from the orbits of close binary stars [75]. Stellar mergers or near-mergers could result in a collimated outflow. However, the binary merger scenario cannot explain the existence of multiple pairs of lobes in different orientations. Since binary star mergers cannot be recurrent, they are unlikely to be the cause of episodic mass ejections.

An exception, however, can be found in the case of CK Vul, which provides an example of multipolarity seen in a merger event. It had a nova-like event in 1670–1672 [76], now thought to have been a stellar merger involving one degenerate component, which ejected a large bipolar nebula [77,78] with velocities ~2000 km s$^{-1}$ at multiple polar angles [79]. Galactic stellar mergers tend to show bipolar ejecta [80,81], perhaps not surprising in an event dominated by angular momentum loss. In contrast, extragalactic mergers are assumed to be spherically symmetric, an assumption well known from planetary nebula history.

Rotating jets are seen in some planetary nebulae and proto-planetary nebulae, with M2-9 being a notable example. Images taken over several epochs have found a corkscrew-like pattern of the outflow [82,83] (Figure 7). These patterns may be caused by wind bending in the motion of a binary system [84–86]. A morpho-kinematic model has been developed for the rotating jet in NGC 6543 [31]. These rotating motions differ from the multipolar morphologies in that the ejection appears to be continuous rather than episodic. Such point-symmetric structures can already be seen in proto-planetary nebulae [87] (Figure 5).

The likely explanation involves a disk, where one star in a binary system experiences accretion and jet formation [88]. There is a striking similarity between the rotating jet of the planetary nebula Fleming 1 (PN G290.5 + 07.9) [89–91] (Figure 8) and the symbiotic star R Aqr, where high-resolution imaging has revealed that the jet comes from one component in a binary system [92,93]. The example of Fleming 1 indicates that this process can continue for some time after the AGB mass loss has ceased. However, a notable difference between the two objects is that R Aqr is a wide single-degenerate binary with a period of 44 years, and Fleming 1 is a close double-degenerate binary with a period of 1.2 days. Similar physics may operate at very different size scales.

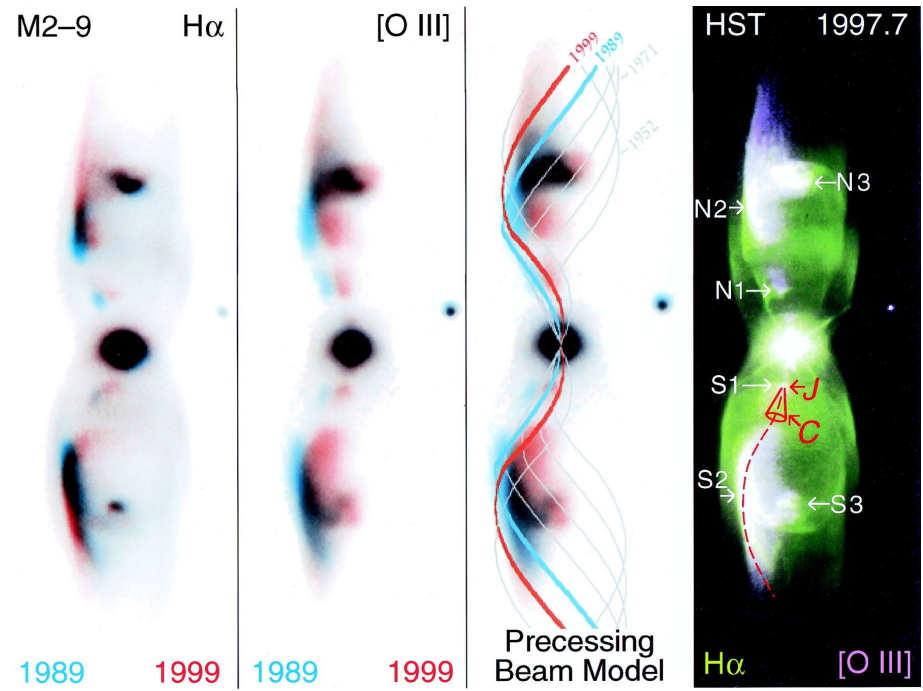

**Figure 7.** Hα images (left) and [O III] (2nd from left) images of M2-9 obtained in 1989 (cyan) and 1999 (red). A rotating beam model is overlaid on the [O III] image on the 3rd panel from the left. Each line is a model in which the projected corkscrew turns by 30°. Right: Unconvolved *HST* Wide Field Planetary Camera 2 images from 1997.7 in Hα (green) and [O III] (magenta). A red "J" marks the point where the southern jet contacts the outer lobe, forming a bright [O III] knot. The cone marked "C" indicates the spray pattern emanating from J. Figure adapted from reference [82].

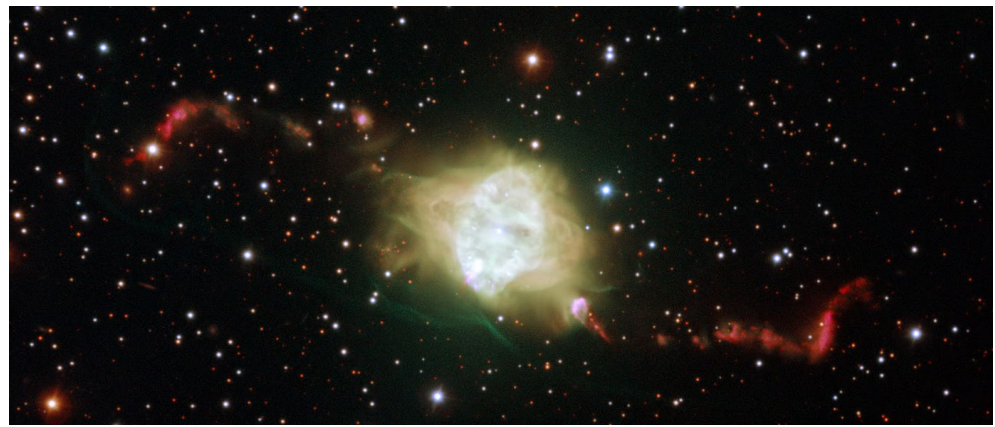

**Figure 8.** A pair of rotating jets can be seen in Fleming 1. This color composite image of Fleming 1 (Hα + [N II] in red, [O III] in green, and [O II] in blue) was taken with the *Very Large Telescope* (*VLT*). Image credit: *European Southern Observatory* and H.M. Boffin.

## 4.3. Episodic Outflows

Series of concentric arcs, separated by nearly equal intervals, have been found in planetary nebulae and proto-planetary nebulae [94–96]. Examples of multiple concentric arcs in NGC 7027 and NGC 6543 are shown in Figure 9. These arcs are probably 3D geometrically thin spherical shells projected onto the sky. Multiple 2D rings are observed in planetary nebulae NGC 6881 [97] and Hb12 [33]. The existence of multiple arcs and rings separated by regular intervals suggests the presence of episodic outflows, either in the AGB phase or in the post-AGB phase. For proto-planetary nebulae, their appearance depends upon the illumination of the central star. The morphology of the arcs and rings could be the result of a bi-conal cavity carved out by fast outflows in multiple spherical shells, as discussed in terms of a 3D model of the Red Rectangle [98].

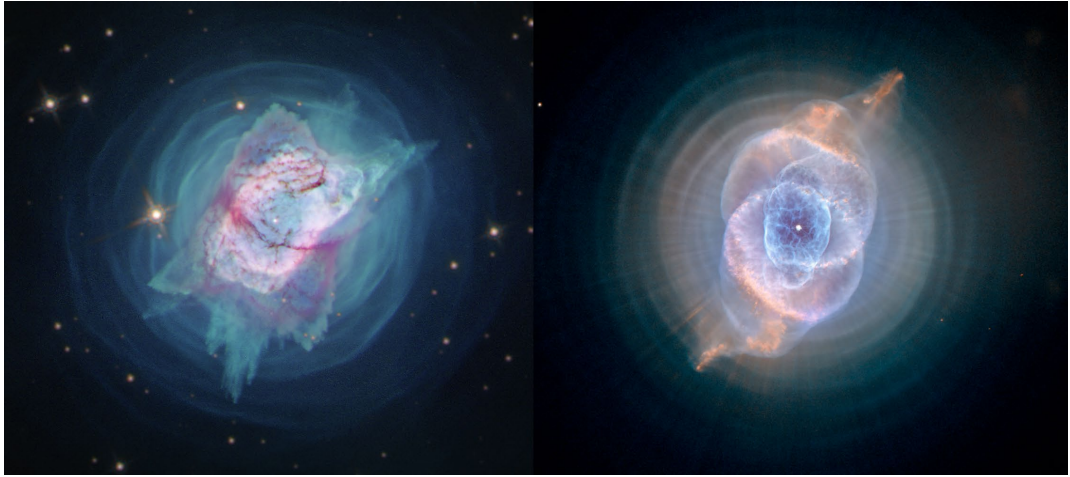

**Figure 9.** A series of concentric arcs can be seen outside of NGC 7027 (**left**) and NGC 6543 (**right**). These color composite images are made from a combination of narrow-band images. The main nebula of NGC 7027 is mainly in white. Credit: *NASA*, *ESA*, and for NGC 7027, Joel Kastner.

*4.4. Microstructures*

The Helix Nebula (NGC 7293) is known for its large number of cometary globules, often with tails pointing away from the star. The globules are seen in $H_2$ and CO emissions [99,100]. Similar globules are seen in NGC 6720 [101], NGC 3132 [102], and NGC 2346 [103]. The globules have diameters of typically a few hundred AU. They occur in similar numbers in each planetary nebula, of around $20–40 \times 10^3$, and contain as much as half the total mass of the nebula [101]. The globules appear to be in pressure equilibrium with the surrounding ionized medium.

The origin of globules remains under discussion. Two possibilities are that they form during the planetary nebula phase, or that they are already present in the AGB wind. Clumps containing a substantial fraction of the envelope mass are seen in AGB winds [104]. Whether these clumps can survive through the transition to the planetary nebulae phase to be responsible for the knots seen in planetary nebulae is unclear, as they are not in pressure equilibrium with the surrounding wind. Notably, all planetary nebulae with such globules are located on the cooling track, where the ionized gas is recombining. This makes it plausible that the clumps result from instabilities at the ionization front, caused by the fact that higher-density regions have much shorter recombination timescales [101].

Knots are also observed in collimated outflows. Figure 10 shows a series of knots on both sides of the outflow of He2-90 [105]. The collimated outflow has a speed of 150–360 km s$^{-1}$ assuming a distance of 1–2.5 kpc. The outflow extends over 40 arcsec, with a kinematic age of ~1400 yr. The knots are ejected at 35–40 yr intervals. Such knots can be produced by a magnetized jet with periodic velocity variations [106].

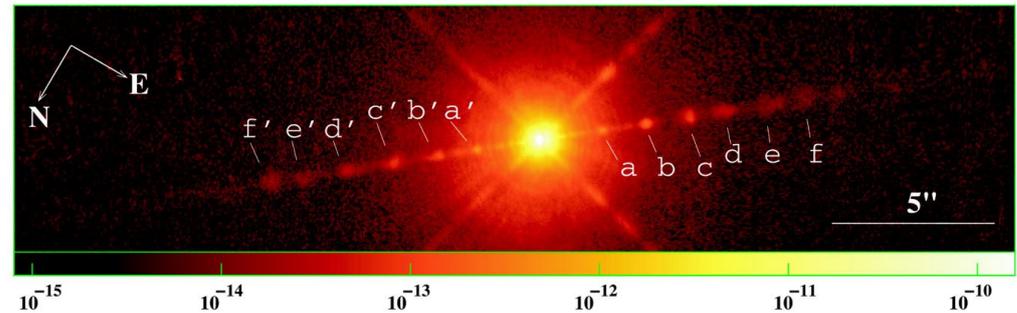

**Figure 10.** Hα (F656N) image of Hen 2-90 taken with *HST* WFPC2 (shown in logarithmic scale) showing a series of knots (labeled as *a*, *b*, *c*...) along the bipolar jet. The very bright central source results in strong diffraction spikes at ±45° in the image due to support structures in the telescope. Figure adapted from reference [105].

### 4.5. Spirals and Disks

Spiral structures are seen around some AGB stars [107]. Figure 11 shows the example of spiral patterns seen in scattered light in the envelope of AGB star AFGL 3068 [108]. These spiral structures are different from the concentric arcs and rings discussed in Section 4.3. A possible origin of the formation of the spirals is through orbital motion, where the center of mass loss is moving along the orbit. There are two effects that contribute to the creation of the spiral structure: the changing location of the star and the changing velocity vector of the star. Compared to the pinwheels seen around massive stellar systems, the AGB spirals tend to be more broken up in arcs. They survive if the wind speed on either side varies by more than the sound speed. Otherwise, the compressed gas expands and dissolves. This causes the spiral to become fragmented.

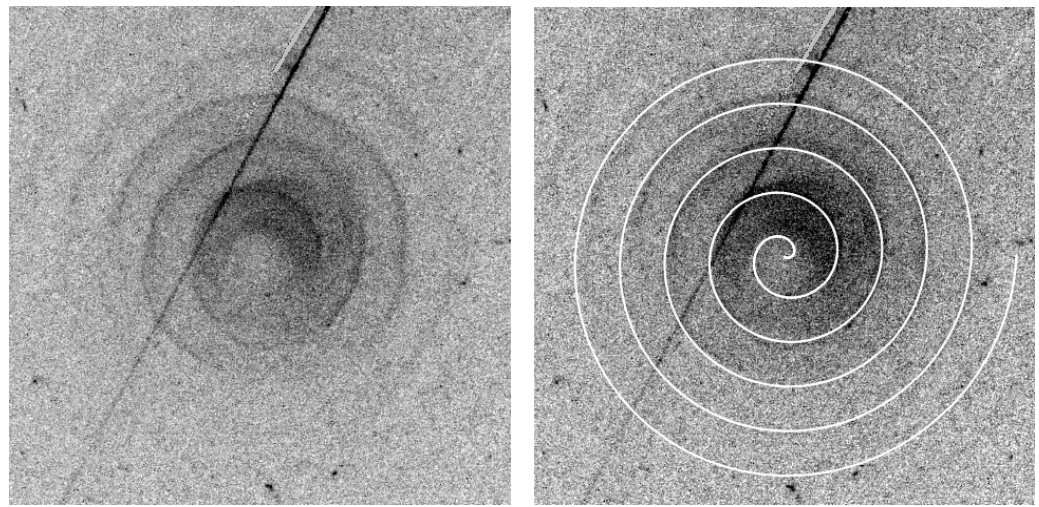

**Figure 11. Left**: *HST* ACS F606W image of AFGL3068 showing a spiral pattern that can be traced to 12 arcsec from the star. The image is due to scattered light from dust in the envelope. The bright straight line refers to a diffraction spike of a star in the field of view. **Right**: Same image with an Archimedes spiral fit. Figure adapted from reference [108].

Planetary nebulae also show spirals, though not as commonly as their AGB progenitors. Ionization of the gas and the increase in the sound speed tend to destroy the pattern, but they may survive in the halo. Examples are seen in the *James Webb Space Telescope* (*JWST*) images of NGC 3132 [102] and NGC 6720 [101]. Orbital periods derived from these spirals are typically a few hundred years. They provide circumstantial evidence for binary companions, which are difficult to detect photometrically. The spiral structure may have some effect on the morphological evolution of the planetary nebulae [109].

A fundamental parameter in the development of the spiral structure is the ratio of the momentum in the companion star to that of the wind it is interacting with. A higher value of this ratio (a more massive companion and/or a tighter orbit) causes torus formation, while a lower value will result in the formation of a spiral [57].

Compact dust disks are found around the central stars of some planetary nebulae. They have sizes of order 100 AU and masses that can range from a large planet to a small asteroid (10% of the mass of Ceres) [102,110,111]. The best observed case is that of NGC 3132 [102,112] where the disk is spatially resolved, with possible evidence for a gap in the disk. The origin of these disks is not known. They may be residuals from AGB mass loss or remnants of debris disks. Evidence for an evaporating gas disk in a young planetary nebulae would favor a mass-loss origin of such disks [113], although different disks may have different origins.

Modern high-dynamic-range imaging has revealed many new morphological features of planetary nebulae, as illustrated in the subsections above. While these features are likely the result of interacting winds, the time-dependent, direction-dependent nature of winds creates a rich variety of morphologies. Currently, the physical origin of the directional outflows and the collimating mechanisms are unknown.

## 5. Dust and Molecular Components of Planetary Nebulae

### 5.1. Molecular Torus as a Collimation Agent

While most of our perception of planetary nebulae is based on optical images, which arise from emission lines from the ionized component, most of the mass of planetary nebulae in fact resides in the molecular component. The distribution of molecular gas can be traced by mapping of molecular emission by mm/submm interferometers. For example, the optical bipolar lobes of planetary nebulae are often confined at the waist by a molecular torus, as seen in NGC 6302 (Figure 12). Recent mappings of six bipolar planetary nebulae with *ALMA* also show that the molecular gas is confined to the equatorial torus [114].

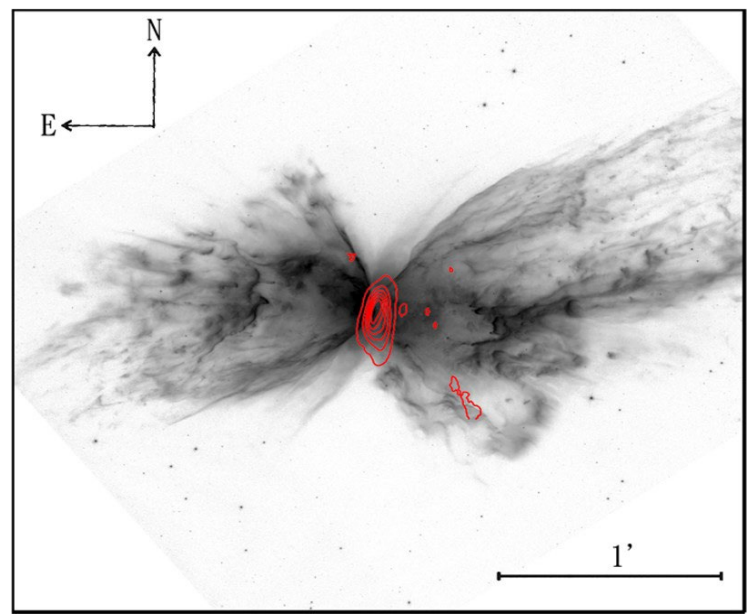

**Figure 12.** Distribution of CO emission (in contours) as observed by the *Submillimeter Array* (*SMA*) [115] overlaid on an optical image of NGC 6302.

The role of accretion disks in collimating outflows in planetary nebulae has been extensively studied in planetary nebulae literature [116,117]. However, the origin of such disks is uncertain.

The interaction between ionized matter and molecular gas is an interesting physics problem in planetary nebulae. While CO emission traces the distribution of molecular matter, the 2.12 μm molecular hydrogen line traces the photodissociation region [118].

The possible role of the molecular torus in collimating the bipolar outflow is illustrated in the schematic diagram of NGC 6302 (Figure 13). Such a molecular torus can also be traced by infrared imaging of the warm dust mixed with the molecular gas [119].

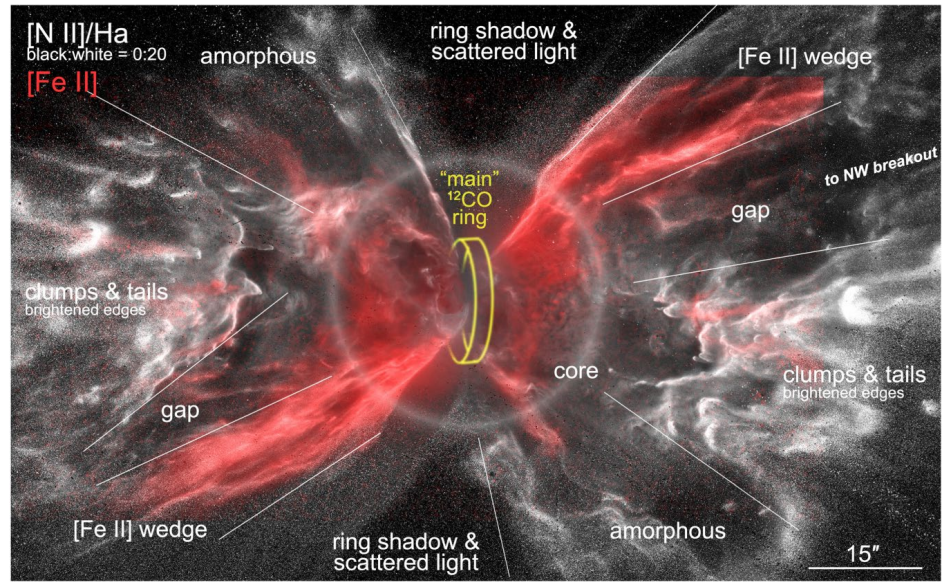

**Figure 13.** A schematic illustration of various features of NGC 6302, including the molecular ring overlaid on a *HST* image of the nebula. Figure adapted from reference [120].

## 5.2. Confinement of Optical Lobes by Neutral Envelope

The optical lobes of many planetary nebulae show sharp boundaries, suggesting that they are confined externally by yet unseen matter. This raises the possibility that the optical lobes represent cavities carved out of extensive molecular envelopes by fast outflows and illuminated by photoionization [121]. Evidence for such large neutral envelopes can be seen in the far-infrared imaging of the bipolar nebula NGC 2346. While mid-infrared imaging shows the torus around the waist of the lobes, far-infrared imaging reveals an extended spherical component outside the optical nebula (Figure 14).

Since molecular emissions arising from the large neutral envelopes of planetary nebulae have low temperatures and low surface brightness, they are difficult to map with interferometers. The cold dust component also radiates in the far infrared, and the mapping of their distribution is beyond the sensitivity of current far-infrared telescopes. An accurate determination of the distribution of the neutral component of planetary nebulae is essential in our understanding of the actual mass distribution of planetary nebulae.

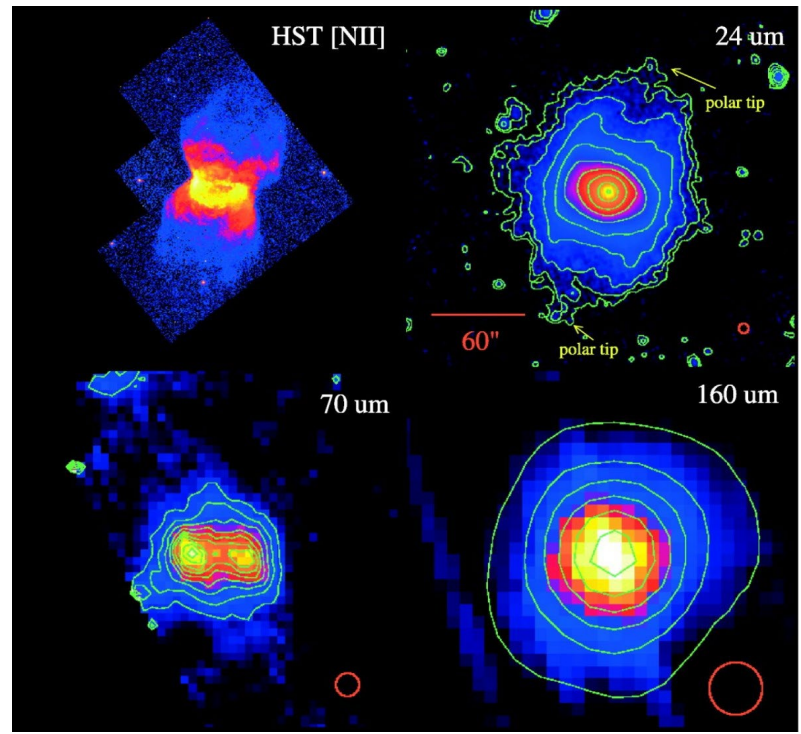

**Figure 14.** Infrared images of NGC 2346 as observed by the Multiband Imaging Photometer for Spitzer (MIPS) of the *Spitzer Space Telescope*. The FWHM beam sizes are shown as red circles in each band. Figure adapted from reference [119].

### 5.3. Circumstellar Chemistry

Many molecular species have been identified in planetary nebulae by mm/submm spectroscopic observations [122]. While some species are formed in the wind of the AGB progenitors, some are synthesized during the post-AGB evolution. Spectral signatures of minerals, including silicates and silicon carbide, are identified in planetary nebulae [123]. The strong unidentified infrared emission bands observed in many planetary nebulae, attributed to vibrational modes of organic compounds, are the result of chemical synthesis during the post-AGB evolution [124]. How such complex organic compounds can be formed in the low-density circumstellar environment remains a major unsolved problem in modern astrochemistry [125].

We now recognize that ionized gas only represents a small fraction of the total mass of planetary nebulae, and most of the mass is in the form of neutral matter. While the ionized component can be easily imaged, the distributions of molecular gas and dust in planetary nebulae are yet to be well determined. A complete mapping of the neutral component is crucial in understanding the role of the neutral component in collimating and confining outflows.

## 6. Large-Scale Structures of Planetary Nebulae

Major advances in understanding large-scale structures of planetary nebulae have resulted from deep, narrow-band and wide-field imagery over large swathes of the galactic plane. Recent Hα surveys of the galactic plane, e.g., the *SuperCOSMOS Hα Survey (SHS)* in the South [126], Isaac Newton telescope Photometric Hα Survey (IPHAS) in the North [127] and the VST Photometric Hα Survey of the Southern Galactic Plane and Bulge (VPHAS+) [128], have led to an explosion of planetary nebulae discoveries [129,130]. While early planetary nebulae catalogs contain mostly high-surface-brightness nebulae, recent deeper surveys have discovered more evolved and lower-surface-brightness planetary nebulae, as well as more compact and younger (and/or distant) examples.

### 6.1. Outer Haloes

There has also been a growth in extremely deep, narrow-band imaging of planetary nebulae undertaken by amateur astronomers [131]. They use small aperture, dedicated telescopes equipped with narrow-band H$\alpha$, [O III] and even [S II] filters that can look at individual objects for tens or even hundreds of hours. These dedicated, amateur programs now regularly provide the best imagery that exists for many planetary nebulae, even of well-known, well-studied examples [132]. These studies also detected some extremely faint central stars that are beyond *Gaia* limits. Such deep imaging can reveal large-scale external structures and low-surface-brightness features such as the halo of NGC 6543 [22]. We should note, however, that external haloes and other structures are not detected in some planetary nebulae regardless of how deep the imaging goes. Clearly different evolutionary processes are at play between those with and without detectable outer structures.

### 6.2. Large Bipolar Nebulae

Another example of large-scale structures is the class of extremely large bipolar nebulae with very high expansion velocities. The extensive outflows in this extreme group can reach several parsecs, and they show expansion velocities of several hundred km s$^{-1}$ and kinetic energies in the order of $10^{46-47}$ erg. These energies are similar to those found in luminous red novae, also associated with intermediate luminosity optical transients [133]. The mechanisms that lead to these extreme bipolar nebulae are uncertain. A possible route of evolution is that these nebulae may descend from post-common-envelope cores, where the close binary was effectively able to successfully expel the entire envelope, adding gravitational energy to the stellar wind and radiation pressure to the mass outflow. Furthermore, if the close pair reaches coalescence it would lead to a violent mass-loss event that would produce a luminous red nova, as is believed to have happened in the case of CK Vul [78,79].

Examples of objects in this group are the bipolar nebulae KjPn 8, He 2–111, and Ou 4. KjPn 8 has three pairs of giant bipolar lobes that cover a remarkable 14′ × 4′ [134]. The distance to KjPn 8 has been derived from the expansion proper motions to be $1.8 \pm 0.3$ kpc [135]. The projected linear extents of the giant lobes are 2.1, 2.4 and 7.4 pc. The bright nebular core was resolved by the *HST* into an elliptical 5.2″ × 2.7″ ring that resembles a young regular planetary nebula [136]. This ionized ring expands at only 16 km s$^{-1}$. A massive, neutral molecular gas disk of ~20″ diameter surrounds the core and expands at 28 km s$^{-1}$ [137]. Within the CO disk, an excited molecular hydrogen disk is detected [138]. The large bipolar lobes expand radially at ~40 km s$^{-1}$ perpendicular to the axes and ~160 km s$^{-1}$ along them. A second, apparently younger, bipolar structure is tilted with respect to the large bipolar lobes. Here, outflow velocities reach 320 km s$^{-1}$ [139]. The kinetic energy of these fast outflows is measured as ~$10^{47}$ erg, which is compatible with an intermediate-luminosity optical transient. The scale sizes and expansion-proper-motion measurements indicate that the giant bipolar lobes were created by three distinct, short-lived events along different orientations over a long period of time. The first bipolar ejection has an age of ~$5 \times 10^4$ years. The middle set of lobes has an kinematic age of ~7200 years and the youngest lobes have ages of 3200 years [135].

The southern nebula He 2-111 (Figure 15) has been classified as an extraordinary type I bipolar planetary nebula due to its highly enhanced nebular nitrogen and helium content [140]. Its central core is formed by a tilted ring with radial filaments, reminiscent of the core of the bipolar nebula, NGC 6302, where expansion velocities of nearly 600 km s$^{-1}$ and kinetic energies of $10^{46}$ erg have been estimated in its outflows [120]. The core of He 2-111 is surrounded by a giant bipolar halo oriented about PA = 130° and covers about 10 arcmin in the sky along the main axis. The distance to He 2-111 has been estimated at 2100 pc. At this distance the linear size of the giant bipolar lobes becomes approximately 6.1 pc. The heliocentric radial velocities in the lobes, relative to the systemic, have been measured in the range of −380 to +360 km s⁻¹, with the fastest knots assumed to have traveled at speeds of 600 km s⁻¹ [72]. The gas in the lobes is ionized by shocks. For an estimated ejected mass of 0.002 M$_\odot$ for all the knots in the lobes, the kinetic energy is ~$10^{46}$ erg. This value is very similar to those found for KjPn 8 and CK Vul. Given the extraordinary characteristics of He 2-111, it has been suggested that it must have originated as a result of a nova-like event [72] or an explosive common-envelope ejection in a close binary system some 8000 years ago[140].

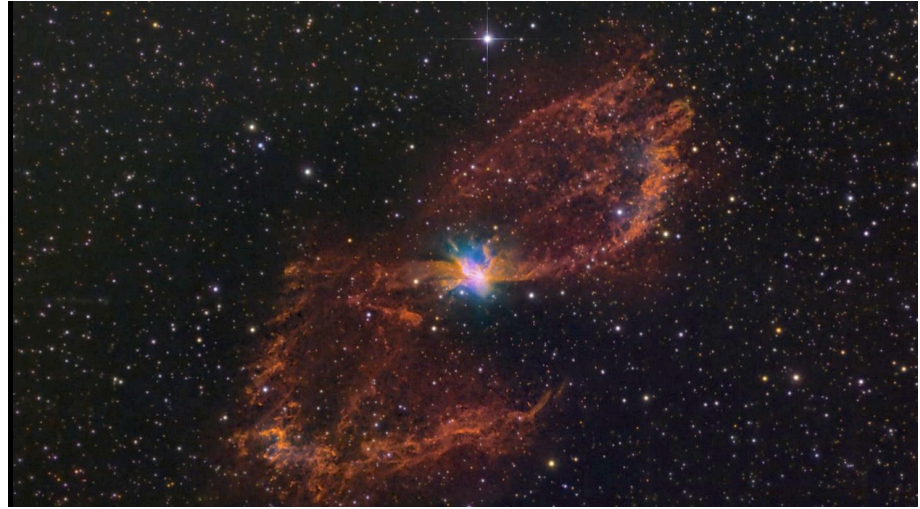

**Figure 15.** Color composite image of He 2-111 (Hα, [O III] and blue continuum). The bipolar lobes extend over 10 arcmin in the sky. Image adapted from reference[140].

Ou 4 is the bipolar nebula with total length of 1.2°, the largest angular extent ever found among bipolar nebulae [141] (Figure 16). It is remarkable that over this gigantic extent the lobes remain well collimated. The central region contains a deformed bubble and the overall large-scale morphology resembles KjPn 8. The nebula is faint and emits mostly in [O III]. Its nature as a planetary nebula has not been confirmed. The distance to Ou 4 is not known with certainty, though it seems to be located inside the Sh 2-129 H II region [142]. The central section of Ou 4 coincides with the triple-star system HR 8119, located at some 700 pc. It is therefore possible that this object is related to a massive outflow from a young, high-mass star. Alternatively, the object could be much closer and descend from a violent post-AGB event, perhaps related to a common-envelope close binary final process.

The core of Ou 4 is of a high excitation class, indicative of a very hot central source. However, as in the case of He 2-111 and KjPn 8, the tips of the lobes are so far away from the exciting central source that they are shock-excited instead of photoionized. The total kinetic energy of Ou 4 is estimated to be ~4 × $10^{47}$ erg [142].

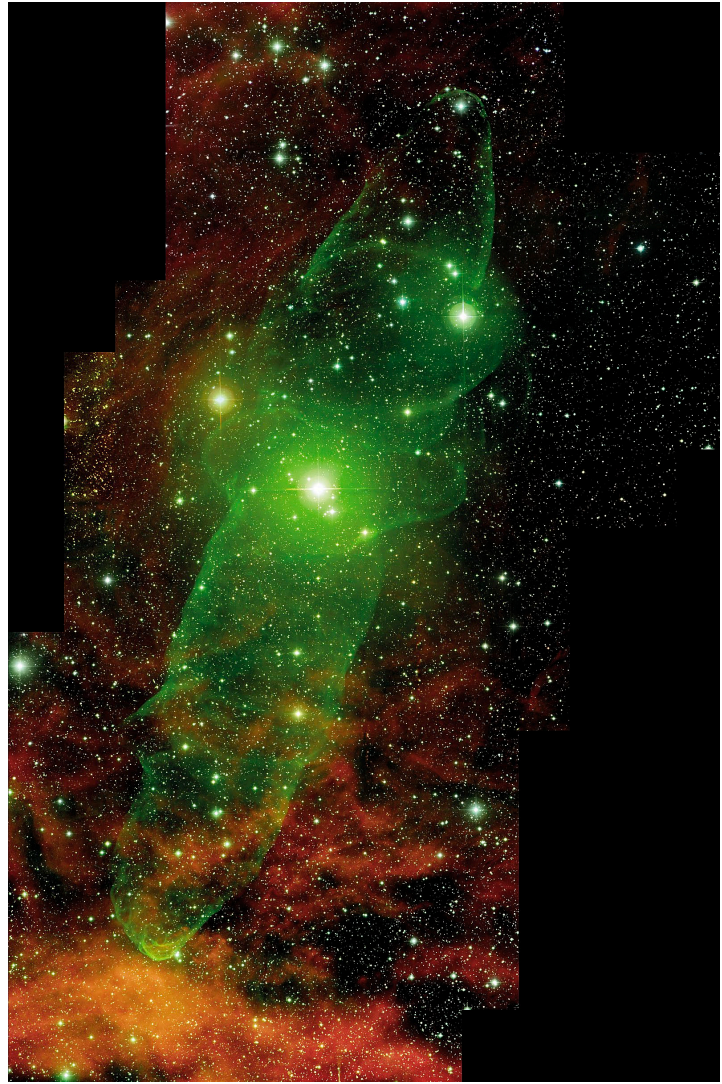

**Figure 16.** Color composite image of the large bipolar nebula Ou 4 ([O III] in green, Hα + [N II] in red, *g* in blue). Image adapted from reference [142].

### 6.3. Large Outer Structures

Wide-field imaging can also reveal previously unseen large structures surrounding planetary nebulae. Large bipolar lobes extending over 100 arcsec have been found outside of the main nebula of IPHAS PN-1 [143]. A large, extended structure has also been discovered outside the optical image of the planetary nebula M1-41 [144].

The existence of large-scale structures in planetary nebulae is unexpected and the physical mechanisms responsible for the large dimensions and hypersonic expansion velocities are not known. Future wide-field narrow-band imaging may further reveal such structures. One possible origin of these structures is eruptive events similar to luminous red novae or intermediate luminosity optical transients [145], where a highly efficient post-common-envelope process expels the envelope or leads to coalescence of the close binary, ending in a violent mass-loss event. The frequency of occurrence of such events is uncertain. Monitoring of the sky for such events will become possible with the Vera Rubin telescope.

## 7. The Relevance of Binarity to PN Evolution and Morphology

There has been a long-standing interest in searching for binary companions around central stars of planetary nebulae. Early photometric monitoring of the central stars of planetary nebulae revealed some binary companions [146], but these numbers were small. Many more were found using the data from photometric imaging variability surveys (OGLE [147]; Kepler/K2 [148]). An early estimate is that at least 20% of planetary nebulae have a binary central star, while some estimates run as high as 50% or more [149]. A recent work puts the fraction of planetary nebulae with binary nuclei between 23 and 36% [21]. The vast majority of those detected are of short period ($P$), less than 1 day [150]. A radial velocity program to monitor central stars of planetary nebulae was initiated and initial results showed some variability [151]. However, the study was not continued long enough to distinguish whether the variability was due to periodic binary motion or variable winds in planetary nebulae central stars. More recently, a few longer-period binaries have been found from sustained radial velocity monitoring (e.g., LoTr5: $P$ = 2700 d [152]; NGC 1514: $P$ = 3300 d [153]).

Thus we presently have an interesting situation in which at least 20% of planetary nebulae appear to have a binary central star and in almost all of these the binary companion is very close to the evolved central star ($P < 1$ day). For close binary central stars of planetary nebulae, there is evidence for good alignment of the axis of the binary orbit with the orientation of the nebulae, strongly indicating that the binary companion shaped the nebula [154,155]. The shortness of these periods, and consequently the closeness of these orbits, indicates that at some point in the evolution, the companion was inside the atmosphere of the evolved star. This has led to an extensive study of the interaction of the two stars within a common envelope, and the role that the energy released in the in-spiral would have in the ejection of the envelope. Most studies find this occurring on a dynamic timescale, on the order of a few years or less [156].

Given that common-envelope evolution is an interesting aspect of cataclysmic variables [157], it has been suggested that this may be the common way in which planetary nebulae form [158–160]. This is in contrast to the standard picture of planetary nebulae formation by the gradual evolution of a post-AGB star and the commensurate expansion of its surrounding nebula, as cited in the Introduction.

## 7.1. Search for Binary Central Stars in Proto-Planetary Nebulae

Might there be an evolutionary link between the central stars of proto-planetary nebulae and the close binary central stars of planetary nebulae? Are the central stars of proto-planetary nebulae the precursors of the close binary central stars of planetary nebulae? A radial velocity monitoring program was carried out from 1991 to 1995 and 2008 to 2022 to search for binarity in the seven brightest proto-planetary nebulae that can be observed from mid-latitudes in the northern hemisphere [161,162]. With brightness of 7–10 mag, they could be observed with small 1–2 m telescopes, for which monitoring time was more readily available. Proto-planetary nebulae, with their F–G spectral-type stars, have more and narrower spectral lines than in the hotter planetary nebulae central stars and thus can be observed with higher precision. All were found to vary, with periods of 35–135 d, the same as the periods found in their light curves. However, these were shown to be due to pulsations of the stars rather than binarity [163,164]. No firm evidence was found for longer or shorter periods that might be due to binary motions. This is the case even though the morphologies of the seven nebulae are elongated, bipolar, or even multipolar.

These studies allowed one to set constraints on any undetected binaries; if they are present, they must have periods longer than 30 yr or secondary companions with masses less than 0.2 M⊙. This initial sample was chosen based simply on the apparent brightness of the central stars. To increase the probability of finding binary central stars, one might choose a sample with characteristics favoring binarity, such as strong bipolarity in its morphology or evidence of a longer photometric period in its light curve, which could be caused by periodic obscuration due to the barycentric motion of the evolved central star within a circumbinary disk. The latter characteristic has been found to be associated with binarity in the RV Tau variable objects with long secondary periods (class RVb), where periods of 600–2600 d have been found in the light curves and the binarity confirmed by radial velocity studies [165,166].

Recent light curve studies of proto-planetary nebulae have identified six proto-planetary nebulae with long photometric periods, five of them ranging from five to ten years and the sixth being 19 yrs [167,168]. A radial velocity study of one of these, IRAS 08005–2356, has been completed and confirmed a binary period of 7.3 yr [169]. Radial velocity monitoring studies of the others are presently being undertaken to search for binarity. Hopefully this will lead to more clarity on the presence of binary components in proto-planetary nebulae. If these recently identified proto-planetary nebulae with long photometric periods are indeed binaries, then such objects are the likely progenitors of the long-period binary planetary nebulae central stars.

Strong bipolarity is epitomized by bipolar nebulae with tight waists, and those seen nearly edge-on would project the full velocity motion. However, in these the central star is consequently obscured in visible light. If the opacity is not too high, the star may be seen in the near infrared. This appears to be the case for IRAS 17245–3951, and Hrivnak and collaborators obtained some initial observations that suggested that the star might vary in velocity. However, this was not pursued due to the lack of instrumentation and the difficulty in getting observing time on 8 m class telescopes.

Thus there is no direct evidence of a close binary companion in any of the proto-planetary nebulae. Also, the common-envelope phase through which the close binary central stars are thought to pass has a timescale of a few years or less. By contrast, a few dozen proto-planetary nebulae have been observed for over 30 years without showing the drastic changes in their properties (brightness, color, spectral type) expected for stars passing through the common-envelope phase with the ejection of their envelopes. Nor is there evidence, such as rapid rotation, to support the idea that the companion has merged with the evolved central star. Rather than evidence for evolution on a short, dynamic timescale, the expansion ages of their nebulae appear to be a few hundred to a few thousand years, timescales that agrees with post-AGB evolution models [170]. Preliminary evidence for the thermal evolution of two proto-planetary nebulae over the course of a decade at a rate in agreement with that of stellar evolutionary models of single post-AGB stars seems to support their identification as single stars evolving to the planetary nebula stage [171].

So, we seem to be left with the following unanswered questions: Where are the binary companions thought to be the drivers of the shaping of the nebulae in proto-planetary nebulae? Are they low-mass (<0.2 $M_\odot$) stars or even brown dwarfs or planets, or do they have long (>30 yr) periods? Where are the progenitors of the close binary central stars of planetary nebulae? They do not appear to be the identified proto-planetary nebulae. Are there two different ways, gradual post-AGB evolution and common-envelope evolution, for planetary nebulae to form from AGB stars?

There are hundreds of branches of binary star evolution, depending on the initial masses of the two components and the binary separations [172,173]. If one or more of these branches leads to planetary nebulae, the exact pathway needs to be identified.

### 7.2. Hierarchical Systems

There is growing evidence for the presence of more than two stars within some planetary nebulae [174,175]. NGC 3132 and NGC 6720 both have known binary companions, but *JWST* data provides indirect evidence for a further companion in each system [101,102]. Triples are also being found for AGB stars [176].

The fraction of stars with binary companions in the solar neighborhood ranges from 30% for 1 $M_\odot$ stars to 80% for 4 $M_\odot$. The fraction of triples ranges from 10% to 30% over this mass range [177]. Of the main sequence stars with close binary companions, 40% contain a third star. Such systems form hierarchical triples, with two stars on a close orbit and a third star much further out. Compared to these numbers, the fraction of binarity among planetary nebulae seems lower than expected, rather than higher! And the effect of hierarchical triples on their evolution cannot be ignored.

There are many uncertainties about how hierarchical triples evolve, and in what state they reach the AGB and planetary nebulae phase [178]. It has been found that the outer star causes the inner pair to harden [179]. In 40% of wide white dwarf–white dwarf binaries, the more massive star is a merger product. Triple systems are the dominant channel for stellar collisions [178]. The outer component of a hierarchical triple may often become unbound during AGB mass loss, but it will still be close to the inner system during the planetary nebulae phase, as is seen, e.g., in NGC 6720 [31]. Models of the impact of a triple system on the outflow and morphology are being developed, for wide outer systems [179] and for tighter outer systems [180].

### 7.3. The Abundance Discrepancy Problem and Possible Connection to Binarity

The emission lines of planetary nebulae are important tracers for elemental abundances [181]. Some refractory elements are highly depleted because of dust condensation [182,183], while other volatile elements retain their stellar abundances in the gas phase. Abundances are mostly measured from the strong forbidden lines; however, they can also be determined from the much weaker recombination lines. The two sets of abundance determinations are often not in agreement, with the recombination lines yielding higher abundances. This discrepancy can reach extreme levels, an order of magnitude or more in some cases. As the forbidden lines are temperature-sensitive, the difference is commonly attributed to a cool, hydrogen-poor component of the nebula [184], but the nature and location of this component is unclear. In some cases there could be a relation to the recombining high-density knots found in high-mass planetary nebulae on the cooling track [101], but this is not confirmed and it is unlikely to be a universal explanation.

Wesson et al. [185] confirmed a strong relation between extreme abundance discrepancy factors (ADFs) and the presence of a close binary system, with orbital periods on the order of a day. These are post-common-envelope systems. But it is unclear how common-envelope evolution could cause the ADF anomaly. A possibility is that it relates to turbulence in the nebula. The orbital velocities are high and may impart significant turbulence on the mass loss during the spiral-in phase.

## 8. Planetary Nebulae in Extragalactic Systems

Tens of thousands of putative planetary nebulae have been observed in systems outside our galaxy, typically using on–off [O III] band imaging. These planetary nebulae candidates have been used for distance determinations, determination of the Hubble constant, tracers of galactic dynamics, estimation of the baryonic mass, and for determination of dark-matter distribution. A surprising result from the extragalactic planetary nebulae studies is the planetary nebula luminosity function (PNLF). It shows a cut-off at the high luminosities, which seems largely invariant to the age of the underlying stellar population. Models of stellar evolution have had difficulty explaining this invariance: they predict more planetary nebulae at luminosities above the cut-off, and a lower cut-off in older systems. The faster evolution shown by the new post-AGB stellar evolution models [170] has resolved part of the discrepancies [186] but not all [187]. Old stellar populations where star formation fully ceased more than 5 Gyr ago are still predicted to have fainter planetary nebulae, unless a low level of star formation continued for longer. The deficit of observed high-luminosity planetary nebulae may be due to the effects of internal dust extinction [21,187]. Binary interactions [188] are less likely to be the cause of the invariance, as they do not produce fixed end products of the interaction. The PNLF is a powerful tool for constraining the evolution of planetary nebulae.

As we discussed in Section 2, the strength of the [O III] line alone is not a reliable indicator of planetary nebulae status. While many of these candidates will be true planetary nebulae, many will also be mimics based on our studies of galactic planetary nebulae. Since most of the extragalactic planetary nebulae are unresolved, we are unable to use their morphological characteristics or physical sizes to verify their true planetary nebulae status. However, the Magellanic Clouds are close enough that we can use imaging and spectroscopy to perform more reliable assessments and analyses. For example, some planetary nebulae candidates found in the *Dark Energy Camera Magellanic Clouds Emission-line Survey* have been found to be H II regions [189]. Some bright nebulae are too large to be planetary nebulae and too small to be massive star bubbles. Direct comparisons of planetary nebulae populations in systems with different formation, kinematic, and chemical histories can then be made.

## 9. Summary

Research on planetary nebulae has spanned over two hundred years. While significant progress has been made on our understanding of the origin and evolution of planetary nebulae in the scheme of stellar evolution, recent space-based observations and observations beyond the visible region have discovered new morphological features such as arcs, rings, spirals, and multiple bipolar flows, which require the introduction of new physical processes into our models. The existence of multipolar nebulae is probably the result of collimated fast outflows penetrating the pre-existing AGB circumstellar envelope, but the mechanism responsible for producing such outflows is unclear. If the 2D concentric rings are the result of episodic outflows, what determines the period of the episodes? The presence of secondary bipolar lobes along directions different from the primary lobes suggests that episodic outflows could also change directions. Does the presence of a binary central star affect some of the above processes? And while binarity is commonly found in the central stars of planetary nebulae, it has thus far remained elusive in proto-planetary nebulae. Thus, there is not good evidence for the evolution of proto-planetary nebulae into planetary nebulae with close binary central stars. If not the proto-planetary nebulae, what are their progenitors? Large field-of-view imaging reveals the presence of large outer structures beyond the main nebula. What can these outer structures tell us about the mass-loss history of the star? The evolution from cool AGB stars to an environment heated by ultraviolet photons also introduces new chemical processes that lead to the unexpected formation of complex organics. In spite of the long history of the field, planetary nebulae remain an excellent laboratory for the study of physics and chemistry of the interstellar medium.

Most fundamentally, we need a clear definition of the planetary nebula phenomenon so that our studies are not confused with objects in other stages of stellar evolution.

In order to tackle these unsolved problems, we need high-dynamic-range and high-angular-resolution mapping of the ionized, molecular, and dust components of planetary nebulae. The latter two are particularly important as they contain most of the mass in the planetary nebula system. Integral field spectroscopic imaging would provide much-needed information on the kinematics of various outflow components in planetary nebulae. Kinematic mapping of both the ionized and neutral components would allow us to develop a comprehensive true 3D model of planetary nebulae.

The dynamical evolution of planetary nebulae is the result of interactions of time-dependent (and possibly directionally dependent) outflows at different stages of central-star evolution. The identification of the physical mechanisms driving these outflows remains a fascinating topic of future theoretical research.

**Author Contributions:** All authors contributed to the content of this paper. The final editing was performed by S.K. All authors have read and agreed to the published version of the manuscript.

**Funding:** This research was funded by the GR009036 from the Natural Sciences and Engineering Research Council of Canada. NSTC 114-2112-M-110-004 from the National Science and Technology Council of Taiwan. UNAM-PAPIIT IN111124. the Hong Kong Research Grants Council for GRF research support under grants 17304024, 17326116277 and 17300417. the Jet Propulsion Laboratory, California Institute of Technology, under a contract with NASA (80NM0018D0004), and partially funded by multiple JWST and HST GO awards administered by the STScI under NASA contract NAS5-03127.

**Data Availability Statement:** No new data were created or analyzed in this study. Data sharing is not applicable to this article.

**Acknowledgments:** We dedicate this paper to the memory of our friend and colleague Detlef Schönberner, who contributed greatly to our modern understanding of the planetary nebula phenomenon. This article represents a partial summary of the discussions that took place at the University of British Columbia, Vancouver, Canada, March 2025.

**Conflicts of Interest:** The authors declare no conflicts of interest.